# Exploration of Knitted Spacer Fabrics using Grasshopper

Suzanne Oude Hengel[1] and Loe Feijs[2]

**Abstract**

We describe a novel approach for exploring, visualizing and modelling of complex textile structures. We show this approach in action for the exploration of weft-knitted spacer fabrics. The modelling tools are readily available and represent a fast-growing approach in communities of generative design. The approach helps solving the problem that spacer fabrics are complex structures which are hard to imagine and to visualize. The explorations turn out insight-full and flexible and are complementary to what can be done with existing tools.

**Keywords:** Spacer fabrics, knitting, 3D modelling, parametric design, visualization,

## Introduction

In the world of knitting there is a traditional big divide between two communities, which are industrial knitting and home knitting. In the next paragraph we shall give more characteristics of each, but the important observation is that, for a variety of reasons, this is not a smooth spectrum, but a spectrum with a somewhat difficult space in the middle.

Industrial knitting is a well-established industry which began in 1589 when the English Reverend William Lee invented the weft-knitting machine with a 16-fold greater productivity than a manual knitter (Schrank et al., 2014). Between 1802 and 1806 Pierre Jeandeau is said to have invented the first latch needle (Hawkins), in 1816 Marc Brunel, arranged the needles in a circular form (Hawkins), and in 1878 Griswold added a second set of needles to the top of circular knitting machine (Hawkins). The American Civil War (1861-1865) and the 1$^{st}$ World War (1914-1918) gave enormous boosts to industrial knitting (Hawkins).

The global knitted fabrics market reached a value of nearly $55.8 billion in 2018, having grown at a compound annual growth rate of 2.1% since 2014, and is expected to grow at 4.6% to nearly $66.9 billion by 2022 (Research and Markets 2019). The knitted fabrics form the second largest market segment of fabrics (31.7%) coming after woven fabrics (45.7%) (Research and Markets 2019).

These fabrics are used to make garments for a global market which is huge: the Global Knitwear market is anticipated to reach USD 699 Billion, growing at a rate of over 5% due to increasing e-commerce and expanding fashion industry (TechSci, 2021). Moreover, rising awareness about health and increased participation in activities, such as running and yoga, are elevating the demand for active wear. Based on the product type, the market has been segmented into Innerwear, T-Shirts & Shirts, Sweaters & Jackets, Sweatshirts & Hoodies, Shorts & Trousers, Evening Dresses, Suits, and Leggings & Accessories. In 2019, t-shirts, shirts and innerwear acquired more than 40% share which is anticipated to increase in the future (TechSci, 2021).

Industrial knitting relies heavily on the knitting machine industry, which produces the sophisticated knitting machines based on two centuries of development. The market for industrial machines is dominated by a few heritage brands from Germany and Japan, notably Stoll (Germany) and Shima Seiki (Japan), both for flat knitting, Karl Mayer (Germany), Fukuhara (Japan), Monarch (USA), and Santoni (Italy). Many smaller companies are active from Spain, Taiwan, China, and so on. This industry tends to protect its intellectual properties by patents, for example Espacenet gives 764 hits for knitting and Stoll, 759 for knitting and Karl Mayer (5 overlaps as Karl Mayer has bought Stoll in 2020), 2.469 for knitting and Shima Seiki. The market for knitting machines was 6.8 billion in 2018 and is expected to grow to 10.8 billion in 2027 (Credence, 2021).

[1] Suzanne Oude Hengel, Arnhem, The Netherlands

[2] **Corresponding author**: Loe Feijs, Studio LAURENTIUS LAB., Sittard, The Netherlands, and Department of Industrial Design, Technische Universiteit Eindhoven (TU/e), The Netherlands, l.m.g.feijs@tue.nl

In the last decades, computer technology has become an essential part of knitting design and industrial production: there are embedded microcontrollers inside the machines, controlling all moving parts, and there is CAD software for designing the knitted fabric structure, the color patterns, and the panel layouts. Most file formats are proprietary and the software is expensive. Besides the above-mentioned dominant heritage brands, there are many small companies active in knitting, witnessed by thousands of patents addressing specialized applications such as shoes (for example)

Home knitting has always been widely popular and still is today. With just a few simple knitting needles, almost any garment can be knitted, which means that anyone can begin learning and practicing knitting without initial investment (except for an investment in time). There are also affordable knitting machines for home usage, although these are sold in much lower numbers than home sewing machines ─ many Brother machines are still around, although Brother stopped producing them in the late 1990s, and there are Chinese clones (BrotherWiki). Nowadays, there is a shift in purpose however. Until one or two generations ago, knitting was a necessity for low income families to have warm garments for the cold season. During long winter evenings, as outdoor work was difficult, the women would knit socks, gloves, sweaters, vests, and hats for their husband, children and grandchildren.

Nowadays, the main purpose is creative expression (BrotherWiki), making special garments with special patterns: attractive items which are much more personal than the ready-mades from industry. For many types of products, it isn't a matter of cost anymore, it is often cheaper to buy a mass manufactured product. Techniques and patterns are widespread and readily available from friends, knitting magazines, websites and *Youtube* movies. The community website *Ravelry* has about 9 million registered users, and approximately 1 million monthly active users. Sustainability and social considerations play a role too, for example working with wool from local farmers, well-cared sheep, recycling of yarn, and so on.

Between industrial knitting and home knitting there is a creative space with lots of possibilities but also lots of difficulties. It is a space populated by specialized artists, exceptional amateurs, museums, graduation students, and tiny companies. The possibilities are endless: any structure, any form, no matter how small or how large can be knitted. There are connections to the fab-lab movement; indeed, there is an abundance of small-scale projects bringing home-knitting machines under Arduino control, combining knitwear with electronics, in the same physical spaces and communities where also laser-cutters, 3D printers, semi-automated looms and so on are combined. But the complexity of knitting itself, the time-consuming nature of manual or semi-automated knitting, and the proprietary file formats make it difficult to connect to the power of the industrial machines. It is in this in-between space that the project we report in this paper took place.

The carrier for our exploration was designing spacer fabrics, but similar situations and opportunities arise for other knitted structures. The difficulty of knitted spacer fabrics is that they need a double-bed machine in order to get high-quality results, and there is no way to prototype the construction using hand needles, for example. In an industrial production setting, samples would be made, mechanical properties measured in a dedicated lab, and after a lot of (expensive) exploration, the structure would go into (mass) production, the initial lab costs being recovered by the profit of significant turnovers. In our case, as in most fab-labs, the driving force is curiosity: can we make this? Can we push the limits further? Can we connect to other creatives? This is where we spotted an opportunity to import a modelling approach from the world of 3D printing, where it is central to "generative design". In particular, we use Grasshopper, a 3D programming language, which is quite unique in the landscape of CAD tools and programming languages. We use it to explore and understand the structure of our knitted spacer fabrics, for which otherwise it is very hard to see what is going on inside and almost impossible to explain to others.

## What is Grasshopper?

Grasshopper (www.grasshopper3d.com) is a graphical programming language for building 3D structures. Unlike most other programming languages such as Java, Javascript, Matlab, Mathematica, and Python, the program logic is described by connecting basic building blocks in an interactive and graphical way. A similar way of programming was pioneered in MAX/MSP before, which is popular among music composers (Manzo, 2016). In Grasshopper, each building block performs a simple task, for example replicating an object multiple times, translating or rotating an object, drawing a line between two points, and so on. There are also sliders so it becomes easy to play with parameters. This feature enables a design approach known as "parametric design" or "generative design". The language was developed by David Rutten working at Robert McNeel & Associates. Grasshopper is not stand-alone, it is a plug-in for Rhino, a traditional 3D CAD tool (Becker and Golay, 1999). Rhino already had a large user base, mostly in mechanical engineering, *before* Grasshopper was added. Rhino had no built-in notion of material or material properties (Unlike Solid-works), but its objects were defined by surfaces, surfaces in turn being defined by lines, and lines by points. In its core, the main mechanism under the hood of Rhino is the mathematical description of smooth curves called NURBS (Rogers, 2001), somewhat like the popular Bézier curves, but more versatile. Rhino is often used as a what-you-see-is-what-you-get (WYSIWYG) editor of 3D structures. The fortunate circumstance was that Rhino came with a scripting API, which enabled Rutten's step of putting Grasshopper on top of it. Figure 1 shows two windows, one for Rhino and one for Grasshopper, working together.

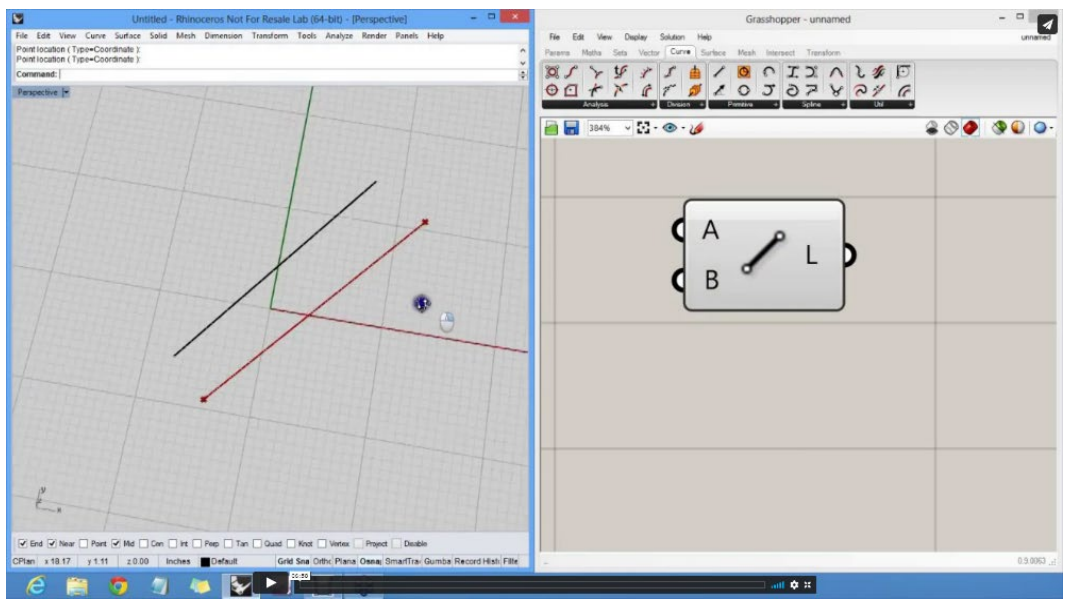

**Figure 1.** *Left is the Rhino window with a coordinate grid and two lines, one made in Rhino in a WYSIWYG manner (black) and one made in Grasshopper (red). Right is the Grasshopper window with one "line" component. Source: Grasshopper Getting Started by David Rutten – 01, Vimeo.*

The user has two windows, to the left in Figure 1 is a classic Rhino 3D model, to the right is a window for the Grasshopper components. The two views are connected in the sense that some of the WYSIWYG-made objects (lines, curves, surfaces, 3D objects) are fed as input to some of the Grasshopper components. It is also possible to see the output of the Grasshopper component network in the same 3D world of Rhino, or add them by a step called "baking".

Grasshopper is deployed in several application domains which vary widely in scale. It is used for city planning (Krishnamurti, 2012), for architecture (Gogolkina, 2018), for product design (Sun, 2019), and for mathematical art (Godthelp, 2019). In case of product design, the production methods are typically 3D printing, CNC milling and laser cutting.

## What are spacer fabrics?

According to (Spencer, 2001) (chapter 30, section 10): "*A spacer fabric is a double-faced fabric knitted on a double needle bar machine. The distance between the two surfaces is retained after compression by the resilience of the pile yarn (usually mono-filament) that passes between them.*" Typical applications include body protection, mattresses, sportswear, car interiors, seats, and shoe soles. Spacer fabrics are an alternative to rubber, foam and natural filler materials with potential advantages of being lighter, more permeable for air, less prone to retain moisture, more environment-friendly, and more stable over time. We show the main idea in Figure 2.

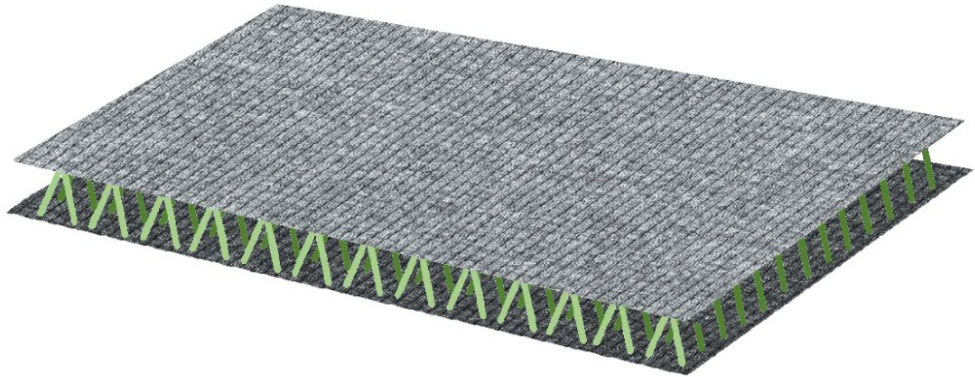

**Figure 2.** *Sketch of spacer fabric with upper and lower face and pile yarns ("spacers") in between.*

Spacer fabrics have been an active area of invention for about 15 years. The plot in Figure 3 shows the number of patent applications on the query "spacer" and "fabric" in Espacenet (own research, January 2021).

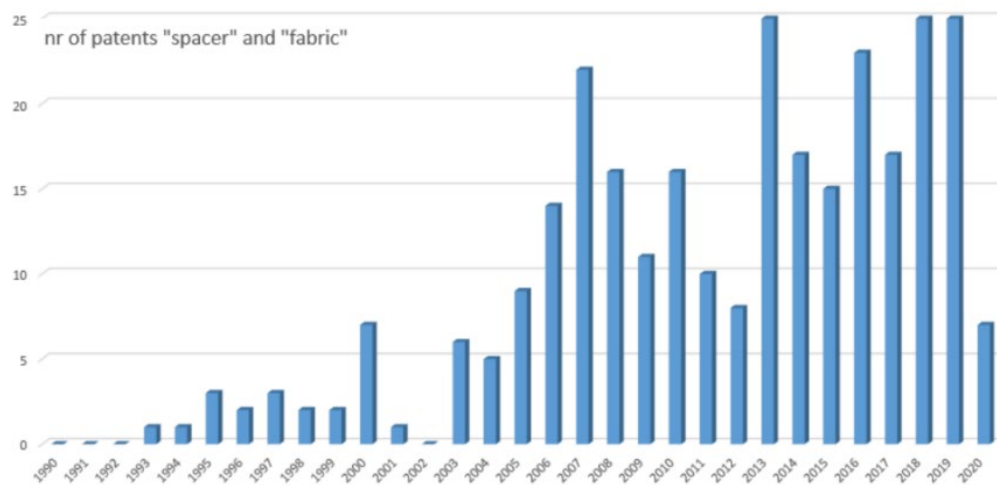

**Figure 3.** *Spacer fabric patents from 1990-2020.*

As for all of knitting, there are two main types of structures (Spencer, Chapter 6): warp knitting (usually on circular machines) (Spencer, Chapter 7) and weft knitting (either by hand or on flat bed machines) (Spencer, Chapter 23). Here we work with weft knitting, but similar work could be done for warp knitting as well.

The design of spacer fabrics is difficult for several reasons. One reason is that it is hard to see what is going on during production, because of the high-speed of the automated machines, and the small scale of the knit structure itself. Unlike traditional knitted fabrics, the most interesting part of the wafer fabric is inside and hard to analyze without cutting into the fabrics (which would change the parameters immediately). There are various dynamic effects such as the stretching of the fabric when inside the machine followed by relaxation once the fabric comes off the needles. And of course there is the dynamic effect during usage that the monofilaments resist, bend or buckle when under pressure, which is what eventually gives rise to the desired (or undesired in case of buckling) transversal elastic properties (Bruer, 2005). Despite the difficulties, the design space is large (it is already large for non-spacer knitting). It's not just the material properties of the pile yarns which matter, but also how they are positioned: vertical, or more skewed? Left-right or also in the longitudinal direction, or both? At which spacing? It should be noted that the position of the pile yarns (also called spacers) cannot be chosen arbitrarily, but has to be aligned with the upper and lower fabric by appropriate stitches. It is a huge puzzle, and for the designer, seeing the "puzzle pieces" in front of his/her very eyes is vital for making progress.

## Approach

One question, which was the starting point of deploying Grasshopper to visualize spacer fabrics, was what would be an effective ratio of the horizontal (left-right) and vertical (longitudinal) spacer distances? There was an intuitive hope to find some kind of simple formula, 4:8 perhaps, or another simple ratio that would work best. This was also the point where the two authors met, one being an experienced creative knitter, the other having a background in design and engineering. Knitting is a very technical subject area with its own language, so one of us had to go through a steep learning curve.

After lots of exploration and discussion on what type of model would be both feasible and useful, a modelling approach emerged, based on the following choices:

1. Level of detail: we abstract from the yarn deformations which happen inside each stitch, yet we do make the stitches visible and we allow for stretching of an entire course or wale of stitches.
2. Programming versus WYSIWYG: we describe the shape of a basic loop, a float stitch and a tuck stitch as a curved line in Rhino; the fabric itself is then described by Grasshopper operations using appropriate components for repetition, counting, translation, and so on.
3. Stretching and raising: the interplay of the fabric stretch and the arising of inter-panel distance, which is the core mechanism why a spacer fabric does make space, should be modeled explicitly in the core of the model. Other features can be added in an on-need manner, once the core is understood.
4. Visualization and animation: we give thickness and color to the lines of the stitches and spacer yarns to make sure that the fabric looks realistic and natural and that thus the entire model becomes a communication tool. All the Rhino view options (Becker and Golay, 1999) are available, so by mouse interaction, the user can pan, zoom and rotate the entire model in a virtual 3D space, which is most helpful to see the structure (much better than frozen renderings). Moreover the key model parameters are obtained from sliders, so the user can play with the model. The sliders can also be connected to a timer, in which case an animation is generated.

In the next sections we show this approach in action and we shall discuss its weak and strong points.

## The core model

We describe the core of our model by showing the Rhino views of one and the same model in two different states, obtained by choosing two different values for the krimp parameter (meaning shrink, which can be manipulated through a slider). There are two layers of weft knitted panels, which we call the upper and the lower panel. There is one spacer yarn, shown yellow, which in reality would be a stiff monofilament, which is connected to both panels by tuck stitches. After each tuck, the spacer yarn floats for two stitches and then it is tucked in the opposite panel, and so on. In practice there will be many more such spacer yarns, but here we just show one, for simplicity. Whereas in Figure 4, the default shrink factor is 1.0, in Figure 5 it is 0.98. Shrinking means releasing the stretch which is naturally built-in to each panel because of the knitting machine's working. In the model this is done by a horizontal scaling of each panel, simply multiplying the inter-stitch distance by 0.98 (which is fed into translation and repetition components of Grasshopper). The tuck stitch is modelled rigid and so is the straight-line segment of the spacer yarn. Only at the connection of the tuck and the straight line, the spacer yarn flexes (in the model).

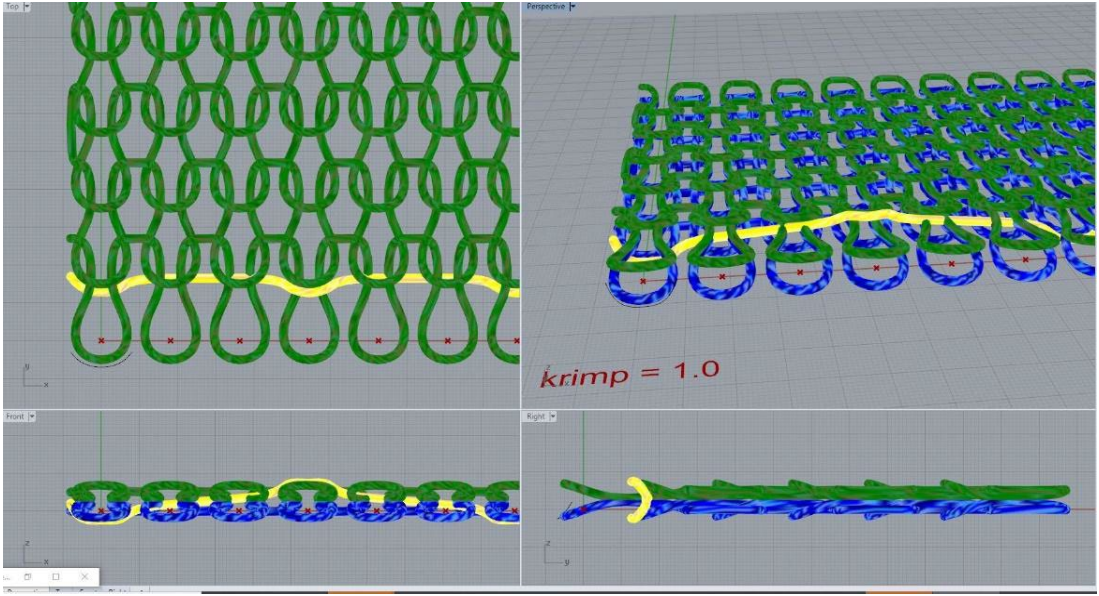

**Figure 4.** *Spacer fabric with one course of horizontal spacers (yellow) with panels (blue, green) stretched.*

This shows the core principle why the panels are moved apart. In physical reality, this happens by the elasticity of the panels and the stiffness of the spacer yarns, which push and pull against each other near the tuck stitches. Other subtle details could be modelled as well, like how the panel stretches by either elasticity in the weft yarns or loop deformations, and all kinds of friction or play.

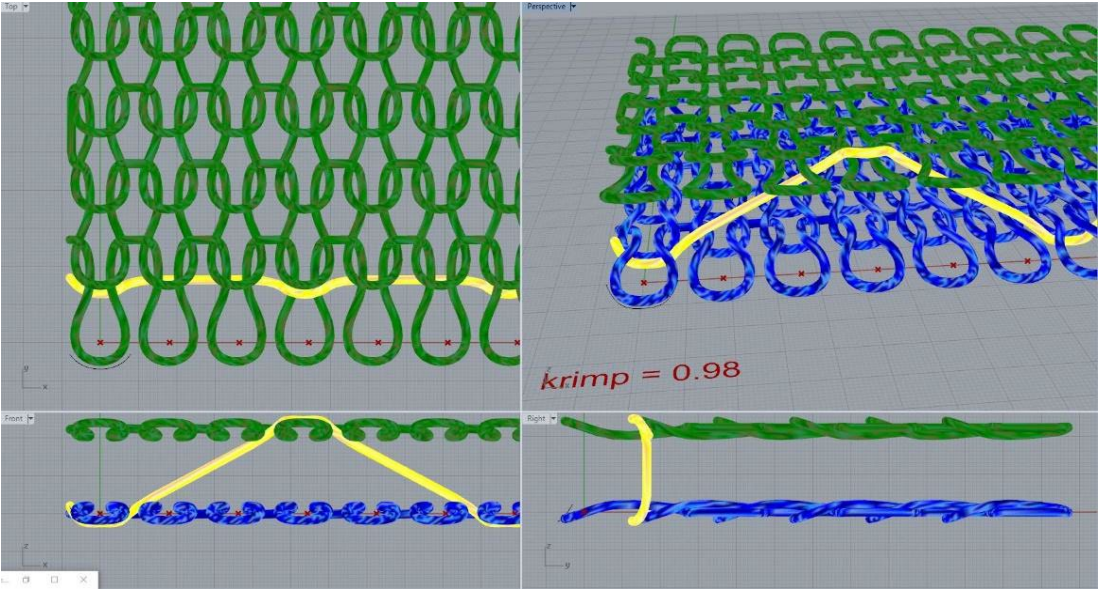

**Figure 5.** *Spacer fabric with one course of horizontal spacers after shrinking of the panel.*

So the space between the panels depends on the horizontal stitch width, the shrink factor, and the length of the spacer yarn floats. The latter length, which in reality is determined during knitting (when the inter-panel distance is minimal), depends on the number of missed stitches of the spacer yarn's float. We considered two possibilities to model this mechanism of panels being pushed apart:

1. To use a special physics engine plug-in to Grasshopper. There is one such engine which is gaining popularity in architecture (e.g. for free-form hanging roofs), called Kangaroo (Piker, 2013). We could deploy it to define joints, constraints, anchors, and elasticities and see the resulting distance emerge from the iterative equations solver inside Kangaroo.
2. To figure out the equations between the key parameters ourselves and transform the equations into a computation. For the latter computation, there would again be two options: a) model all numeric multiplications, additions etc. by Grasshopper components, one for each operator, or b) write them as a formula and embed the formula as a mini-Python program in a single Grasshopper block (indeed, there are many plug-ins, Python being one of them).

We took the second option: once we saw the configurations in early models, similar to Figure 4 and 5, and having played with a few parameter settings it became clear that essentially the distances involved are related by equations based on Pythagoras' law. No iterative solver is needed and we felt it would be beneficial to have simple equations rather than an outcome emerging by magic from a solver. We began with sub-option 2a; later, when working with spacers in two directions, we switched to option 2b.

The equations are explained using Figure 6, which is based on 14 Gauge knitting (Gauge simply refers to the number of stitches a garment has per inch.). Referring to Figure 6, we recognize two rectangular triangles, one with sides $A_i$, $B_i$, and $C_i$ (subscript $i$ for initial), the other with sides $A$, $B$ and $C$. The former describes the initial condition, when the just-knitted fragment of each panel is still on the needles. We have $A_i = m \times H$, and $B_i$ is assumed given as the distance between the two beds of the knitting machine. Then $C_i$ follows by Pythagoras' law $C_i = \sqrt{(A_i)^2 + (B_i)^2}$. When the stitches come off the needles, they shrink by a factor we call σ (for example 0.98 in Figure 5). In the new situation, $A = \sigma \times m \times H$, and $C = C_i$ (because the spacer is considered rigid). Now $B$ follows from Pythagoras, working backwards, and thus we find the inter-panel distance as $B = \sqrt{C^2 - A^2}$.

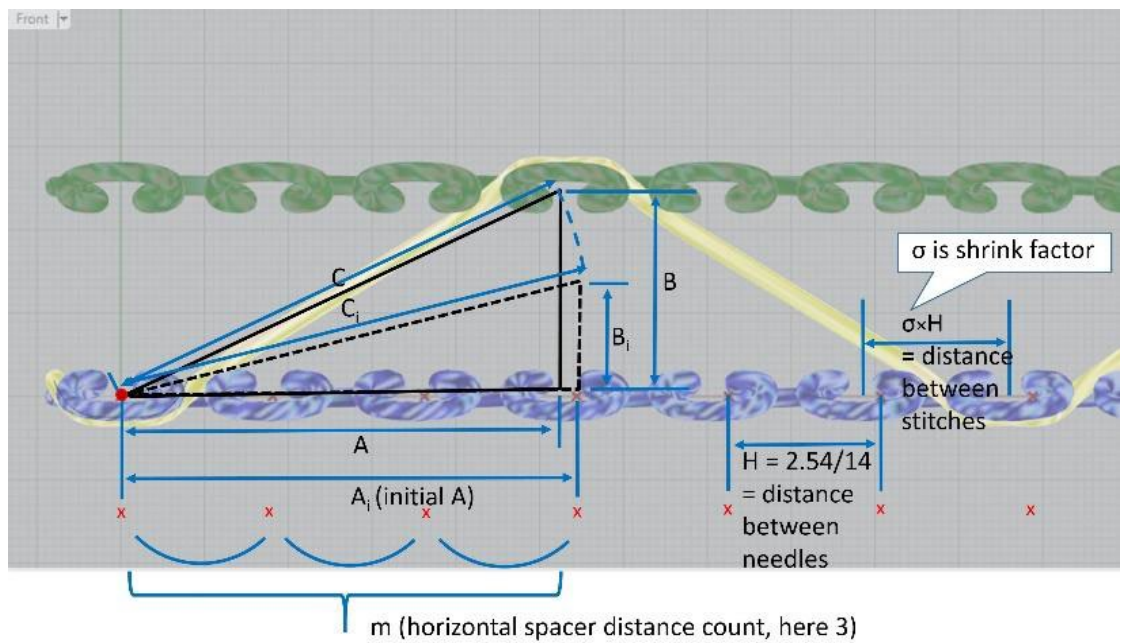

**Figure 6.** *Finding the inter-panel distance B from the other knitting parameters.*

This core model is the basis of more refined studies, where the same style of modelling is used again and again. Although more parameters play a role, the equations stay of the same type: no real new hurdle is encountered. It would be outside the scope of this paper to describe all refinements we did, but we summarize them in one of the next sections (Section *extensions and experiences*).

## Language aspects

It is important to point out that the models, of which Figures 4 and 5 are screenshots, are not hand-crafted. They are the output of a parametric Grasshopper program, which has been assembled as a network of sliders, repetition components, arithmetic components, mesh pipe components, coloring components etc. To give an impression of the complexity of the network, we include Figure 7. It is easy to build-up such a network, connecting each component's inputs to other components' outputs by straightforward drag and drop. It is slightly more difficult, however, to read the program back after a while, or explain it to someone else. That's why we took great care to add comments and group the components such as to suggest a logical flow of the data. Moreover, we adopted a helpful color scheme. The pink zone in Figure 7 is about elementary in-course distances such as the gauge and the shrink factor, whereas the dark grey zone is about inter-course distances. The blue flow is for the lower panel, the green flow for the upper panel; the yellow flow is for the spacer yarn (in other words, we give the program's land areas the same colors as the yarns themselves, i.e. the colors seen in Figures 4, 5, 6).

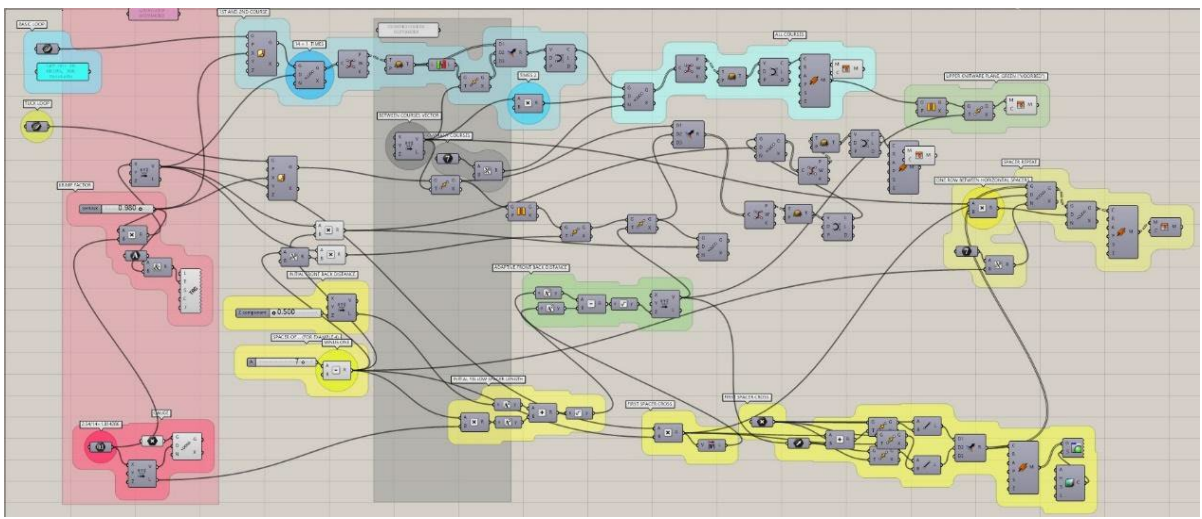

**Figure 7.** *Overview of the core Grasshopper model.*

In an ideal world, this description could be automatically linked to the existing CAD tools for knitting such as provided by Stoll and Shima Seiki, which are needed anyhow to design the stitches and program the machine (in our case a Stoll ADF830-24W flat bed knitting machine and ). The world is far from ideal, however. Most professional knitting software is rather closed, and it is even doubtful whether machine programming and fabric simulation can or should be all in one model (mostly likely it would be too complex to learn and manage). For the time being we work with a patchwork of tools, Grasshopper now being one more patch.

## Extensions and experiences

In addition to the horizontal spacers (aligned along the direction parallel to the courses, like the yellow yarns in Figures 4 and 5), we studied and realized fabrics which also include vertical spacers (in the orthogonal direction, parallel to the wales). From the Grasshopper model we understood that both types of spacers have the effect of pushing the panels outwards, but that they need careful alignment to make sure that they push to the same distance, rather than work against each other (which would weaken the structure). Playing with the models, and building upon the equations given before, we could make the equilibrium condition explicit:

$$\frac{m}{n} = \frac{V}{H} \times \frac{\sqrt{\tau^2 - 1}}{\sqrt{\sigma^2 - 1}}$$

Here, $m$, $H$ and $\sigma$ are the parameters of the horizontal spacers and $n$, $V$ and $\tau$ are the corresponding parameters of the vertical spacers. When the vertical spacers are not fully vertical, because they must shift by one tuck stitch each hop, we get a slightly more complicated equilibrium condition, but in essence, the approach still works. We also explored the tensions in the spacer yarns under non-equilibrium conditions: we animated refined models where the yarns get thinner and show red spots when stretched (Grasshopper allows the tube widths and mesh colors to be defined by computation and thus change dynamically).

Another open problem which was hard to answer otherwise was whether the horizontal and vertical spacers would interfere with each during the relaxation phase, or perhaps would collide for an impossible cross-over. This question could be solved by an animated version of several models with different combinations of $m$ and $n$.

We found that this type of modelling opens the gate to developing a number of insights and possibilities:

- Visualization of the internal structure and better understanding of the internal mechanisms of complex fabrics, such as the equilibrium formula given above. Some of the insights correspond to insights found experimentally by other spacer fabric researchers (see section *related work*), others are new.
- Improved communication with specialists and non-specialists to share and exchange ideas. In particular, screenshots and animation movies were helpful in a project where a knitter (the first author) and a weaver (MV) cooperate with different fabric construction methods to develop innovative spacer fabrics.
- The fact that we had a mathematical model, triggered another research line, viz. to perform stress-displacement tests. Although we have not reached the point where we have agreement between theory and lab-tests, the discussion of quantitative experimental effects not seen in the model (e.g. the negative resistance) is very stimulating for studying more literature and is helpful to refine our language of mechanisms and effects.
- The approach helps to connect the process of knitting innovation to other innovation processes, which happen in typical fab-lab environments, where 3D printing, electronics, laser-cutting, computerized embroidery, programming, etc. are creatively mixed. We envisage combinations of knitting, 3D printing and/or electronics, similar towhere embroidery, 3D printing and electronics are combined (Goudswaard et al., 2020) . Designing such combinations will be facilitated by the shared language and shared models, as Rhino/Grasshopper is used for 3D printing anyhow.

We began harvesting the first, second and third of these two advantages. The fourth is so far only a vision and in that sense it is speculative. We believe however this vision to be very promising, in particular for those who work in the creative space between industrial knitting and home knitting.

## Related work

The mechanism described in Section *the core model* is precisely what is explained by (Liu and Hu, 2011), who conclude that elasticity in the panel yarns causes the panels to be pushed apart by the spacer yarns. As Liu and Hu explain *"The reason is that high shrinkage of the nylon/spandex covered yarn causes the surface loop wales to come more close after the fabric is removed from the machine and this leads to a higher inclination angle of the spacer yarns."* They conclude that the thickness of the spacer fabric is not just a matter of the machine properties, but depends on elasticity of the panel yarns and on a number $NP$ (spacer stitch loop length, similar to our $n$). Lui and Hu: *"So, it is possible to knit spacer fabrics with different thicknesses using elastic yarns for the outer layers on the same flat knitting machine."*

In the stress-displacement test we found areas with negative slope, which is a kind of (differential) negative resistance, also described by (Chen, Hu and Liu, 2015). The core model does not explain that (yet), but our hypothesis is that the monofilament spacer yarns go into an S-form: once in S-form they lose their beam nature and work as springs of lower stiffness. The S-forms (and other types of buckling) could be added to our Grasshopper models using the Kangaroo plugin (Piker, 2013).

(Guo, Long, Sun and Zhao, 2013) develop models similar to ours, but working in Matlab. Like we did, they study the effect of the spacer yarn inclination angle on fabric properties. Their inclination angle is the same as $\arctan(B/A)$ in our model. They study certain combinations of horizontal and vertical spacers but do not derive equilibrium formulas.

## Discussion and conclusions

There is still a lot of work to be done in modelling, testing and verification of the structures and properties of spacer yarns. Our approach has some overlap with other modelers such as (Liu and Hu, 2011), (Chen, Hu and Liu, 2015) and (Guo, Long, Sun and Zhao, 2013). Whereas these researchers are ahead of us in their experimental lab work, we can still claim certain advantages of our approach. The main advantage is the transparency of the models, which gives insightful 3D models parameterized by convenient sliders and easy to manipulate by pan, zoom and rotate options. The Grasshopper models are flexible and easily tunable to a wide variety of research questions. Another advantage is that the Grasshopper community is open; documentation, help and tricks are widely available. The plugin is readily available, and students get free temporary Rhino licences. Our approach helps to open up the world of knitting towards fab-labs and open-source communities and thus foster cooperation beyond the boundaries of knitting.

## Declaration of Conflicting Interests

The author(s) declare no potential conflicts of interest with respect to the research, authorship, and/or publication of this article.

## Funding

The work described in this project was partially done in the *The Space Between* project supported by the WORTH Partnership Project of the European Commission under COSME.

## References

(Hawkins) Mary Hawkins, A Short History of Machine Knitting, *Knitting History Forum*, http://knittinghistory.co.uk/resources (retrieved 1-2-2021)

(Schrank et al .2014) V. Schrank, A. Hehl, K.-P. Weber, ch5 Processes and Machines for Knitwear Production in: Textile technology: an introduction by Thomas Gries, Dieter Veit, Burkhard Wulfhorst (eds), 2014 Elsevier Inc.

(Research and Markets 2019) *Global Knitted Fabrics Market Analysis, 2019: Key Opportunities & Strategies to 2022* - ResearchAndMarkets.com

(TechSci 2021) Tech Sci Research. *Global Knitwear Market By Product Type, By Material Type, By Application, By Consumer Group, By Distribution Channel , By Company and By Geography, Forecast & Opportunities*, www.techsciresearch.com/report/global-knitwear-market/3641.html (summary only). Retrieved 1-2-2021.

(Credence 2021) Credence Report. *Knitting machines market by type, by application; growth, future prospects and competitive analysis.* (summary only) www.credenceresearch.com/report/knitting-machines-market (retrieved 1-2-2021)

(BrotherWiki) *Machine knitting wiki*. Brother. Retrieved 2-2-2021. https://machineknitting.fandom.com/wiki/Brother_main

(Manzo 2016) Manzo, V. J. (2016). *Max/MSP/Jitter for music: A practical guide to developing interactive music systems for education and more*. Oxford University Press.

(Rogers 2001) Rogers, D. F. (2001). *An introduction to NURBS: with historical perspective*. Morgan Kaufmann.

(Krishnamurti 2012) Ramesh Krishnamurti, Tajin Biswas, and Tsung-Hsien Wang. Modeling Water Use for Sustainable Urban Design. In: S. Müller Arisona et al. (Eds.): *Digital Urban Modeling and Simulation*, (Communications in Computer and Information Science 242), pp. 138–155, 2012. Springer-Verlag Berlin Heidelberg.

[Gogolkina 2018] Gogolkina, O. (2018, December). Parametric Architecture in the Formation of Recreational Complexes. In *IOP Conference Series: Materials Science and Engineering* (Vol. 463, No. 2, p. 022066). IOP Publishing.

(Sun2019) Sun, B., & Huang, S. (2019, July). Realizing product serialization by Grasshopper parametric design. In *IOP Conference Series: Materials Science and Engineering* (Vol. 573, No. 1, p. 012078). IOP Publishing.

(Godthelp 2019) Godthelp, T. S., Jorissen, A. J., Habraken, A. P., & Roelofs, R. (2019, October). The timber Reciprocal Frame Designer: free form design to production. In *Proceedings of IASS Annual Symposia* (Vol. 2019, No. 15, pp. 1-8). International Association for Shell and Spatial Structures (IASS).

(Spencer 2001) David J. Spencer. *Knitting technology, a comprehensive handbook and practical guide* (3d edition). Woodhead Publishing Series in Textiles, 2001.

(Spencer Chapter 6) David J. Spencer. Comparison of weft and warp knitting (chapter 6). In: *Knitting technology, a comprehensive handbook and practical guide* (3d edition). Woodhead Publishing Series in Textiles, 2001.

(Spencer Chapter 7) The four primary base weft knitted structures (chapter 7). In: *Knitting technology, a comprehensive handbook and practical guide* (3d edition). Woodhead Publishing Series in Textiles, 2001.

(Spencer Chapter 23) David J. Spencer. Basic warp knitting principles (chapter 23). In: *Knitting technology, a comprehensive handbook and practical guide* (3d edition). Woodhead Publishing Series in Textiles, 2001.

(Bruer2005) Bruer, S. M., Powell, N., & Smith, G. (2005). Three-dimensionally knit spacer fabrics: a review of production techniques and applications. *Journal of Textile and Apparel, Technology and Management*, *4*(4), 1-31.

(Becker and Golay1999) Margaret Becker, Pascal Golay. *Rhino NURBS 3D Modeling.* New Riders (1999).

(Piker 2013) Piker, D. (2013). Kangaroo: form finding with computational physics. *Architectural Design*, *83*(2), 136-137.

(Goudswaard et al. 2020) Goudswaard, M., Abraham, A., Goveia da Rocha, B., Andersen, K., & Liang, R. H. (2020, July). FabriClick: Interweaving Pushbuttons into Fabrics Using 3D Printing and Digital Embroidery. In *Proceedings of the 2020 ACM Designing Interactive Systems Conference* (pp. 379-393).

(Liu and Hu 2011) Liu, Y., & Hu, H. (2011). Compression property and air permeability of weft-knitted spacer fabrics. *The Journal of the Textile Institute*, *102*(4), 366-372.

(Chen, Hu, and Liu 2015). Chen, F., Hu, H., & Liu, Y. (2015). Development of weft-knitted spacer fabrics with negative stiffness effect in a special range of compression displacement. *Textile Research Journal*, *85*(16), 1720-1731.

(Guo, Long, Sun, and Zhao 2013) Guo, X., Long, H., Sun, Y., & Zhao, L. (2013). Theoretical modeling of spacer-yarn arrangement for warp-knitted spacer fabrics and experimental verification. *Textile Research Journal*, *83*(14), 1467-1476.